\documentclass[aps,superscriptaddress,amsmath,amssymb,floatfix,twocolumn,showpacs,amsfonts,longbibliography,notitlepage,superscriptaddress]{revtex4-1}
\usepackage{times}
\usepackage[varg]{txfonts}
\usepackage{textcomp}
\usepackage{graphicx}
\usepackage{subfigure}
\usepackage{color}
\usepackage[colorlinks=true,citecolor=blue,urlcolor=blue,linkcolor=blue,hyperindex]{hyperref}
\usepackage{braket}
\usepackage{float}
\usepackage{overpic}
\usepackage{bm}
\usepackage{array}
\usepackage{multirow}
\usepackage{booktabs}
\usepackage{appendix}
\usepackage{tabularx}

\makeatletter
\newcommand{\fmarki}{*}
\newcommand{\fmarkii}{\ensuremath{\dagger}}
\newcommand{\fmarkiii}{\ensuremath{\ddagger}}
\newcommand{\fmarkiv}{\ensuremath{\mathsection}}
\newcommand{\fmarkv}{\ensuremath{\mathparagraph}}
\newcommand{\fmarkvi}{\ensuremath{\|}}
\newcommand{\fmarkvii}{**}
\newcommand{\fmarkviii}{\ensuremath{\dagger\dagger}}
\newcommand{\fmarkix}{\ensuremath{\ddagger\ddagger}}
                
\def\@fnsymbol#1{{\ifcase#1\or \fmarki\or \fmarkii\or \fmarkiii\or \fmarkiv\or \fmarkv\or \fmarkvi\or \fmarkvii\or \fmarkviii\or \fmarkix \else\@ctrerr\fi}}
\makeatother

\renewcommand{\fmarki}{\ensuremath{\dagger}}
\renewcommand{\fmarkii}{*}
\renewcommand{\fmarkiii}{*}
\renewcommand{\fmarkiv}{*}
\renewcommand{\fmarkv}{x$_5$}
\renewcommand{\fmarkix}{z$_9$}

\begin{document}

\title{Periodicity staircase in a centrosymmetric Fe/Gd magnetic thin film system} 



\makeatletter
\author{Arnab Singh}
\thanks{These two authors contributed equally.}
\affiliation{Advanced Light Source, Lawrence Berkeley National Laboratory, Berkeley, California 94720, USA}
\author{Junli Li}
\thanks{These two authors contributed equally.}
\affiliation{Center for Neutron Science and Technology, School of Physics, Sun Yat-Sen University, Guangzhou 510275, China}
\author{Sergio A. Montoya}
\affiliation{Center for Magnetic Recording Research, University of California San Diego, La Jolla Ca, USA}
\author{Sophie Morley}
\affiliation{Advanced Light Source, Lawrence Berkeley National Laboratory, Berkeley, California 94720, USA}
\author{Peter Fischer}
\affiliation{Materials Science Division, Lawrence Berkeley National Laboratory, Berkeley, California 94720, USA}
\affiliation{Department of Physics, University of California Santa Cruz, Santa Cruz CA, USA}
\author{Steve D. Kevan}
\affiliation{Advanced Light Source, Lawrence Berkeley National Laboratory, Berkeley, California 94720, USA}
\author{Eric E. Fullerton}
\affiliation{Center for Magnetic Recording Research, University of California San Diego, La Jolla Ca, USA}
\author{Dao-Xin Yao}
\email[Corresponding author:]{yaodaox@mail.sysu.edu.cn}
\affiliation{Center for Neutron Science and Technology, School of Physics, Sun Yat-Sen University, Guangzhou 510275, China}
\author{Trinanjan Datta}
\email[Corresponding author:]{tdatta@augusta.edu}
\affiliation{Department of Physics and Biophysics, Augusta University, 1120 15$^{th}$ Street, Augusta, Georgia 30912, USA}
\affiliation{Kavli Institute for Theoretical Physics, University of California, Santa Barbara, California 93106, USA}
\author{Sujoy Roy}
\email[Corresponding author:]{sroy@lbl.gov}
\affiliation{Advanced Light Source, Lawrence Berkeley National Laboratory, Berkeley, California 94720, USA}

\begin{abstract}

Presence of multiple competing periodicities may result in a system to go through states with modulated periodicities, an example of which is the self-similar staircase-like structure called the Devil's staircase. Herein we report on a novel staircase structure of domain periodicity in an amorphous and centrosymmetric Fe/Gd magnetic thin film system wherein the reciprocal space wavevector \textbf{Q} due to the ordered stripe domains does not evolve continuously, rather exhibits a staircase structure. Resonant X-ray scattering experiments show jumps in the periodicity of the stripe domains as a function of an external magnetic field. When resolved in components, the step change along Q$_x$ was found to be an integral multiple of a minimum step height of 7 nm, which resembles closely to the exchange length of the system. Modeling the magnetic texture in the Fe/Gd system as an achiral spin arrangement, we have been able to reproduce the steps in the magnetization using a Landau-Lifshitz spin dynamics calculation. Our results indicate that anisotropy and not the dipolar interaction is the dominant cause for the staircase pattern, thereby revealing the effect of achiral magnetism.

\end{abstract}
\maketitle
{\bf INTRODUCTION} 

The appearance of staircase-like structure is a fascinating phenomenon that is observed in a variety of condensed matter systems. In 2D electron gas, quantized conductance is manifested as a step feature in the Hall effect measurements \cite{QHallEffect}. In quantum materials, interplay of competing interactions with multiple periodicities in a system can give rise to a ground state whose length scales are defined by the modulation of the original periodicities. Example of such modulated periodicities include commensurate and incommensurate phases, such as density waves in solids \cite{Gruner}, stripes and charge density waves in cuprate superconductors \cite{Lee,Tranquada,kivelson}, charge ordered state in manganites \cite{Cheong}, and helical spin structure in magnetic systems \cite{Binz}. A well known staircase structure is the Devil's staircase which appears when a system goes through numerous phase-locked modulated periodicities \cite{Bak,Aubry,Reichhardt}. Devil's staircase have been observed in magnetic systems \cite{Bak,Bak2,Masuda,Fraerman,Kuroda}, liquid crystals \cite{BahrLiqCrystal} and in ferroelectrics \cite{LiSciadv}. Apart from fundamental science the staircase structures have potential technological applications such as in metrology, sensing devices etc \cite{Kiselev2011}. 

Interesting staircase structure in domain size and in magnetoresistance have been observed in Dzyaloshinskii-Moriya interaction (DMI) based solitonic system \cite{PhysRevB.86.214426, PhysRevB.89.014419} Competition between symmetric Heisenberg exchange interaction and the antisymmetric Dzyaloshinskii-Moriya interaction (DMI) can give rise to interesting magnetic textures such as a helix and skyrmion lattice phases \cite{Muhlbauer,Tokura,TogawaSoliton,Bak2}. DMI based chiral magnetic order in a helimagnet is called Dzyaloshinskii type helimagnet structure, while a helical magnetic order due to competition between ferromagnetic and antiferromagnetic exchange interaction is known as Yoshimori-type helimagnetic structure~\cite{Togawa2016}. The chiral magnetic structures in a helimagnet exhibits solitons that can be manipulated by an external magnetic field~\cite{TogawaPhysRevLett.111.197204}. More specifically, the soliton periodicity changes in a step wise manner which is attributed to the discrete changes in the soliton number because of confinement at the grain boundaries \cite{TogawaPhysRevLett.111.197204,TogawaSoliton}. Field evolution of confined helicoids have also been shown to occur via discrete steps in helical magnet MnSi \cite{Monchesky}. The thin film structure of MnSi accommodates a finite number of turns and the jumps are explained due to annihilation of individual turns of the helicoid. 

In this article we report the observation of a staircase-like structure in the field-evolution of the scattering wave vector \textbf{Q} which emerges from the stripe phase of an amorphous and centrosymmetric Fe/Gd system. In contrast to bulk DMI magnets, previously described, the Fe/Gd system is a perpendicular magnetic anisotropy (PMA) system that exhibits dominant dipolar interactions and negligible DMI~\cite{Sergio, JamesLee}. Given the centrosymmetric nature of the Fe/Gd system, magnetic phases will exhibit an equal distribution of opposite helicity magnetic spin textures that on average makes the dipolar magnet globally achiral. We performed resonant coherent soft X-ray scattering to study the evolution of the stripe periodicity as a function of applied perpendicular magnetic fields at various temperatures. Under applied perpendicular fields and temperature conditions it is possible to obtain a hexagonal skyrmion lattice phase that consists of an equal population of left- and right- chirality~\cite{JamesLee}. We observed that the scattering wave vector \textbf{Q} changes in step like fashion with no well defined step height and width. However, when \textbf{Q} is resolved into components Q$_x$ and Q$_y$, the step heights along Q$_x$, were found to be in integer multiple of 7 nm, which is close to the exchange length of the system. At higher temperature the steps were smeared due to thermal fluctuations. 

Our X-ray scattering studies have been complemented by spin dynamics calculations that take into account the net achiral nature of the material. We have simulated an experimentally observed (non-equilibrium) process where a global versus local phenomena delicately balances each other. On one hand the total periodicity of the stripes increase with increasing magnetic field due to an enhancement in the majority stripe width (spins aligned along the field direction). On the other hand this is being counter balanced by the local minority stripe width which cannot fall below a certain size governed by the exchange length of the system. Thus, instead of a bulk macroscopic motion of domain walls over the entire sample, these competing tendencies cause the local forces and energetics experienced by the minority domains to locally annihilate some of these half-periods. This in turn leads to (as observed experimentally and verified theoretically) a local readjustment of the domain sizes. By defining a magnetic domain length scale ratio, $T$, we have {performed a Landau-Lifshitz (LL) simulation to generate steps as a function of the applied magnetic field and show the important role that anisotropy plays in generating the steps in these systems. We have developed a theoretical model for the appearance of steps using exchange interaction, dipole-dipole interaction, anisotropy, and external magnetic field. Our calculations indicate that the origin of the steps lie in the anisotropy term. Even if exchange and dipole interactions are present, absence of anisotropy does not produce steps. Although the appearance of steps do look similar in single crystal DMI material and the amorphous Fe/Gd system under investigation, the physical origin of the steps in the two systems are different.     

{\bf RESULTS} 
 
{\bf Resonant scattering due to stripes}
 
The scattering geometry for the experimental set-up is shown in .~\ref{Fig:scattering}(a). X-ray beam whose energy is tuned to the Fe L$_3$ edge (707 eV) is incident normally on the sample. A pinhole was placed on the beam path upstream 5 mm from the sample to establish transverse coherence of the beam. In this geometry the X-ray photons are sensitive to the magnetization ($m_z$) along the beam direction. The scattering pattern was collected on a charge coupled device camera (CCD) placed about 0.5 m away downstream. Resonant X-ray scattering measurements are sensitive to static magnetic structure ($S(\bf{q})$) and spatial correlation length ($\xi_{s}$). From the position and intensity of the Bragg peaks it is possible to extract information about the periodicity and strength of the magnetic order. Fig.~\ref{Fig:scattering}(b) shows the full field X-ray microscope images (top panel) and X-ray resonant scattering pattern (bottom panel) of the sample. We observed the presence of three distinct magnetic phases, namely, disordered stripe, ordered stripe and skyrmions, obtained by either varying the temperature or applied perpendicular magnetic field. The X-ray microscopy images were obtained by varying the applied magnetic field at 300 K, while the resonant X-ray scattering data was measured  from $L{N_2}$ temperatures to room temperature as a function of the applied magnetic field. In the ordered stripe phase (T = 239K) the domain periodicity (2$\pi$/Q) at remanence is (119 $\pm$ 5) nm. The stripe pattern persists as the field is increased from zero to around 170 mT, when new peaks in the form of a distorted hexagonal pattern start to appear indicating a transition to the skyrmion phase. These observations are consistent with previous findings \cite{JamesLee,Sergio,Singh}.

\begin{figure}[b]
    \includegraphics[width=\linewidth]{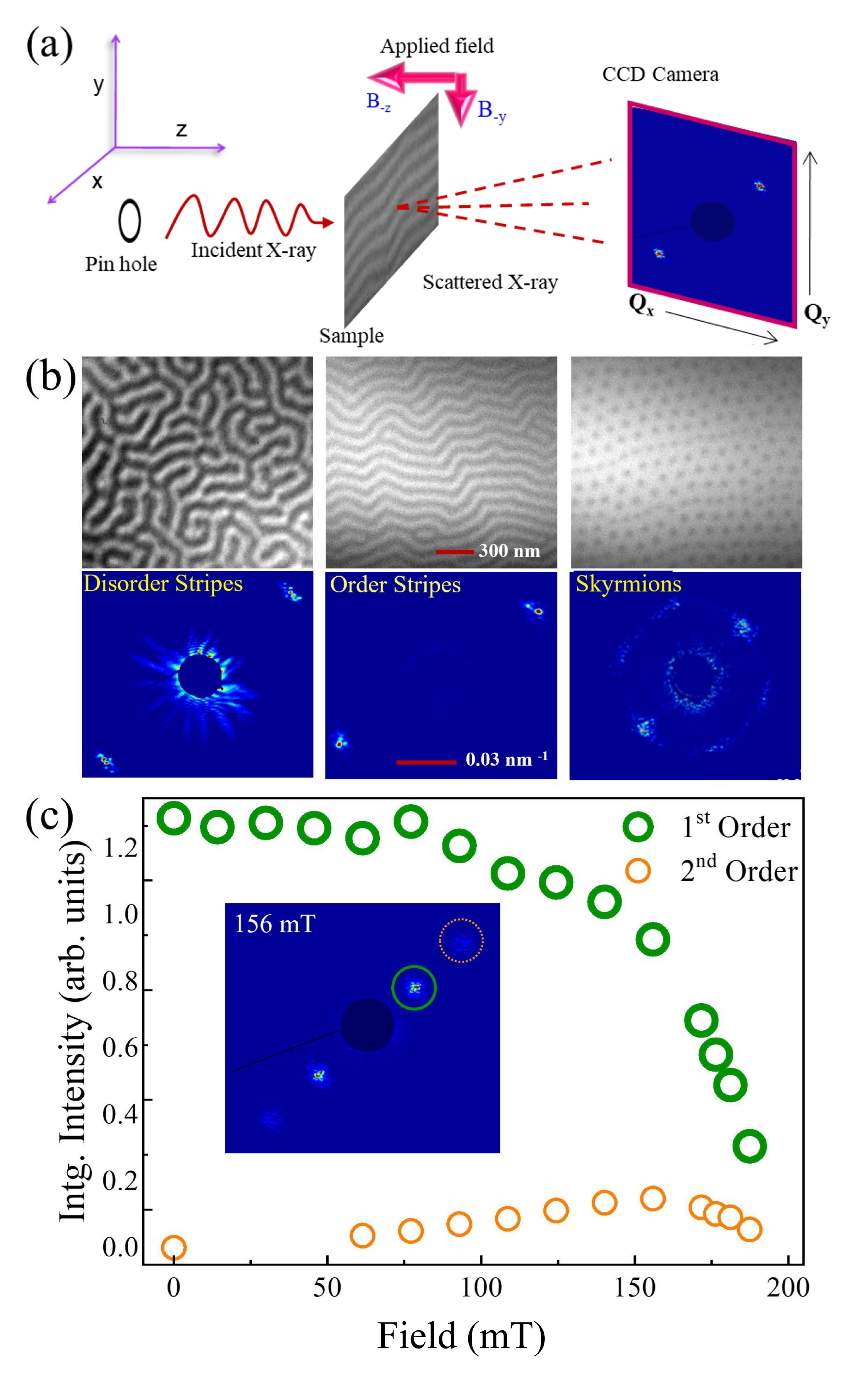}
\caption{{\bf Experimental set-up and magnetic phases.} (a) Schematic of the coherent magnetic X-ray scattering geometry. 
(b) Real space (top panel) and reciprocal space (lower panel) of the different magnetic phases present in a Fe/Gd system. The scattering images were taken at H = 0mT; T = 85K (disorder stripes), H = 0mT;  T = 225K (order stripes) and H = 190mT; T = 239K (skyrmions). A small residual in-plane field is present during ramping down of field from saturation to zero. (c) Variation of the 1st and 2nd order magnetic diffraction peak with field at 239K. Inset image shows the appearance of both 1st and 2nd order diffraction peaks 
}
    \label{Fig:scattering}
\end{figure}

Fig.~\ref{Fig:scattering}(c) shows the field evolution of the integrated intensity of 1$^{st}$ and 2$^{nd}$ order diffraction peak from the ordered stripe domains. 
Starting from the zero magnetic field condition the 1$^{st}$ order peak is a maximum and the 2$^{nd}$ order peak is a minimum ({$\sim$} zero). At remanence the average value of the out-of-plane component of the magnetization is zero. Therefore the widths of the up and down domains are equal. This results in odd-order diffraction spots. Increasing the applied perpendicular field breaks the symmetry between the diffraction peaks, from the stripe phase, causing the even order diffraction peaks to appear~\cite{PhysRevB.74.094437}. Around 170 mT, the intensity of both the 1$^{st}$ and the 2$^{nd}$ order peaks start to diminish and eventually new peaks in the form of a hexagonal ordering pattern appear (see Fig.~\ref{Fig:scattering}(b), bottom right panel). It is interesting to note that in the hexagonal phase we observe two relatively strong intensity spots along the same direction where the stripe peaks are present} which would indicate that somehow the original direction of the stripes is retained even in the hexagonal phase.   
\begin{figure*}[t]
   \includegraphics[width= 16.2 cm]{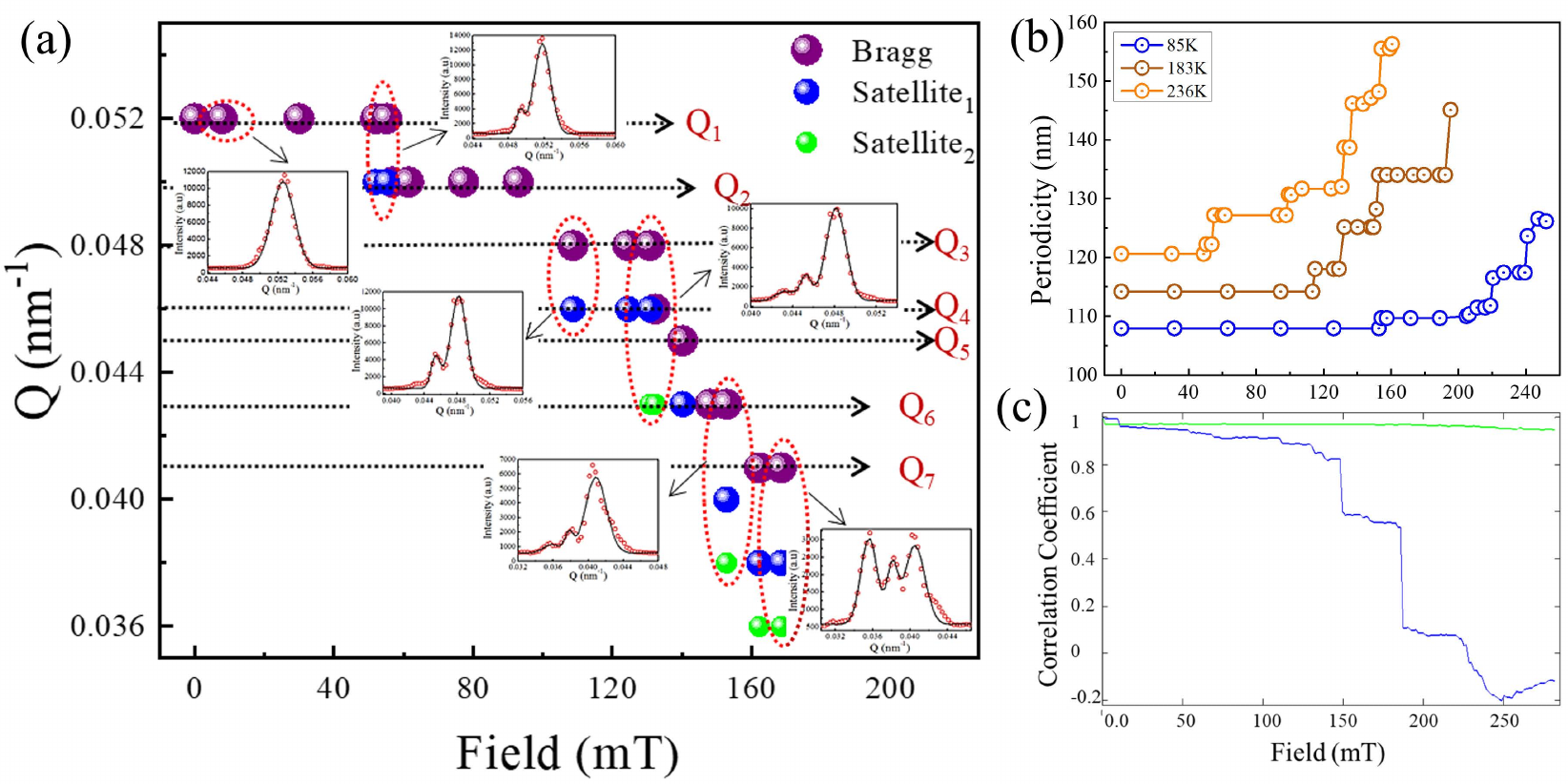}
    \caption{{\bf Evolution of stripe diffraction peak, periodicity and correlation with field.} (a) Plot of the q-vector of the satellite peaks as a function of the applied out-of-plane (OOP) magnetic field at 230 K as the system transitions from magnetic stripe phase to skyrmion phase. Dotted arrows indicate the Q-value of the Bragg peak positions (purple symbol) starting from at different fields. (b) Evolution of the stripe-periodicity with field at various temperatures showing discrete steps like feature. (c) Correlation coefficient values with respect to the remanent state (0 mT) for increasing magnetic field at 85K.}
    \label{Fig:staircase}
\end{figure*}

{\bf Staircase structure of \textbf{Q}-vector}

The evolution of the stripe-diffraction peak in Q-space is shown in Fig.~\ref{Fig:staircase}(a) as a function of the applied perpendicular field at T = 230 K. At the start of the field cycle, the momentum transfer vector, Q$_{1}$ is ( = 0.052 nm$^{-1}$) of the magnetic Bragg peak. As the field increases, the magnetization increases and the size of the favorable perpendicular domains (spins along the field directions) also increases leading to an increased domain periodicity which results in a decrease in the Q-value $2\pi/L_0$, where $L_0$ is the periodicity with respect to zero field Q-value. Interestingly, we observed that the Q-value corresponding to the magnetic Bragg peak decreases in discrete steps as a function of applied magnetic field giving rise to a staircase-like structure. 
 
The evolution of domain periodicity obtained from scattering data happens in several steps that involve sudden jumps and appearance of modulated periodicities. We find that along with the main magnetic Bragg peak, a much weaker satellite peak forms at a smaller Q-value, and both the peaks evolve in an interesting way as the field is changed. The increase in field leads first to the appearance of an initially weaker intensity satellite peak at Q$_{2}$ (at a smaller Q-value than zero field Bragg peak at Q$_{1}$). With further increase in the field, the main Bragg peak (Q$_{1}$) suddenly merges with Q$_{2}$ giving rise to a step-like feature in Fig.~\ref{Fig:staircase}(a). Since the position and intensity of the Bragg peak gives the periodicity of the stripe domains and density of domain scattering respectively, we can conclude that the number of domains with periodicity P$_{1}$= 2$\pi$/Q$_{1}$ decreases with increasing field while the number of domains of periodicity P$_{2}$ = 2$\pi$/Q$_{2}$ starts to increase and finally all domains suddenly transform to the  periodicity P$_{2}$. This sequence of events, changing Q from Q$_{1}$ to Q$_{7}$ with a similar mechanism of peak shifts (Q$_{1}${$\rightarrow$} Q$_{2}$; Q$_{3}${$\rightarrow$}Q$_{4}$; Q$_{5}${$\rightarrow$}Q$_{6}$) was observed throughout the stripe phase (see Fig.~\ref{Fig:staircase}(a)). In some cases, a direct change in Q-values corresponding to the Bragg peaks without any satellite peaks (Q$_{2}$ {$\rightarrow$}Q$_{3}$; Q$_{4}${$\rightarrow$}Q$_{5}$) was also observed.

In Fig.~\ref{Fig:staircase}(b) we convert the wavevector into real space periodicity ($2\pi/Q$) and plot it as a function of applied field at different temperatures. At higher temperatures the total number of steps increases which results in the appearance of the first step at much lower fields for higher temperatures than the lower ones. The plot of the correlation values of the stripe-diffraction spot at different fields with respect to the one at remanence for increasing magnetic fields is shown in  Fig.~\ref{Fig:staircase}(c). Any subtle changes in the speckle pattern between two frames taken at 0 mT and H mT will result in a value of the correlation coefficient (CC) which is defined by

\begin{equation}
\label{eq:correlation coefficient}
\boldsymbol{CC}=\frac{\sum_{m}\sum_{n}(A_{mn}-\bar{A})(B_{mn}-\bar{B})}{\sqrt{(\sum_{m}\sum_{n}(A_{mn}-\bar{A})^2)}{\sqrt{(\sum_{m}\sum_{n}(B_{mn}-\bar{B})^2)}}},
\end{equation}

 where $A$ and $B$ corresponds to the two images taken at two different field values. $A_{mn}$ denotes the intensity value of the pixel position at $m^{th}$ row and $n^{th}$ column of the 2D scattering image. $\bar{A}$ is the mean value of the 2D image. If $CC$ = 1 then the two images are perfectly correlated, $CC$ = 0 means completely de-correlated and $CC$ values lying between 0 and 1 means partially correlated. Thus the variation of the correlation-coefficient can be attributed in the real space as either change in magnetization or density or periodicity of the stripes or any combination of these factors with applied field. 
 
In Fig.~\ref{Fig:staircase}(c) the $CC$ for the stripe phase observed at 85 K is calculated between the scattering image taken at remanence (zero-field) and another image taken at high fields. The field dependent variation of the $CC$ is plotted in Fig.~\ref{Fig:staircase}(c) (blue color line). The correlation coefficient also exhibits distinct steps and most interestingly the horizontal portion of the steps are  populated with small step like features identical to a self-similarity devil's staircase like behaviour. As a measure of the stability of the entire set-up during the measurement we also calculated the correlation coefficients for the Airy pattern, which remains fairly close to unity at all the fields. 

\begin{figure*}[t]
\includegraphics[width= 15 cm]{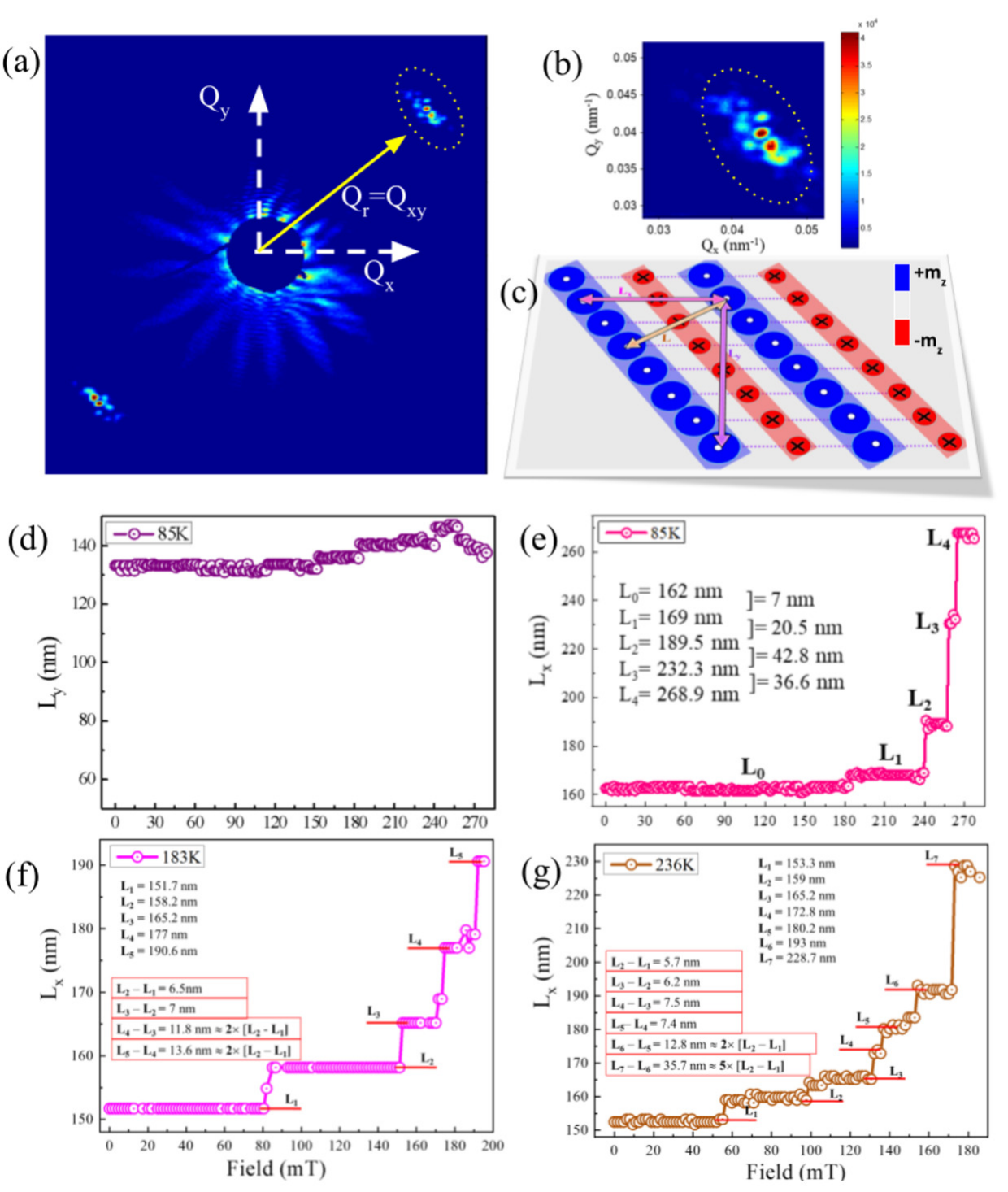}
    \caption{{\bf Stripe orientation and staircase-like behaviour.} (a) A typical scattering pattern of the stripe lattice along with the projection of the in-plane Q-vectors. (b) Enlarged image of the stripe-diffraction spot in Q-space. (c) Schematic real space view of stripe-domain orientation according to scattering image of Fig.~\ref{Fig:staircase}(a), where blue circles with dot resemble the spin along the field direction while the red small circles with cross resemble the spins opposite to the field direction and L= L${_x}{^2}$+L${_y}{^2}$ corresponds to the periodicity of the stripe domains. Plot of the evolution of (d) L$_y$ (= 2$\pi$/Q$_y$) at 85 K and (e-g) L$_x$ (= 2$\pi$/Q$_x$) as a function of the applied magnetic field at T = 85 K, 183 K and 236 K.}
    \label{Fig:QresolvedStaircase}
\end{figure*}

{\bf Resolving staircase along Q$_x$ and Q$_y$ direction }

A typical diffraction pattern consisting of the centrosymmetric first order peaks in the stripe phase is shown in Fig.~\ref{Fig:QresolvedStaircase}(a) along with their in-plane Q-vectors. The enlarged image of the diffraction spot in Fig.~\ref{Fig:QresolvedStaircase}(b) exhibit modulation with speckles indicative of heterogeneity in the ordering of the magnetic domains in the real space. The diffraction spots appears at about ~45$^{\circ}$ to the beam propagation direction (see Fig.~\ref{Fig:QresolvedStaircase}(a)), meaning the stripe-domains are oriented ~45$^{\circ}$ to the X-ray propagation direction. We note here that a small in-plane field of magnitude $\approx$ 1mT is present along y-direction (see Fig 1(a)) in the magnet during rampdown from saturation to zero field. We resolved the resultant \textbf{Q}-vector into Q$_x$ and Q$_y$ components, to get information about the stripe periodicity along real space X and Y direction thereby obtaining real space value L$_x$ and L$_y$ as shown in the schematic representation in Fig.~\ref{Fig:QresolvedStaircase}(c). From this we find that the steps along L$_x$ are significantly distinct compared to that in the  L$_y$ direction (see Fig.~\ref{Fig:QresolvedStaircase}(d and e)). 

Interestingly, we found that the steps in the plot of L$_x$ change in multiples of 7 nm. That is, the minimum change in periodicity along L$_x$ is 7 nm. No such relationship exists for the L$_y$ evolution. The magnitude of the change in L$_y$ is 
typically smaller than 7 nm and random compared to L$_x$ (Fig.~\ref{Fig:QresolvedStaircase}(d and e)). A schematic of a possible stripe domain arrangement is shown in Fig.~\ref{Fig:QresolvedStaircase}(c). The blue (red) domains are majority (minority) domains. The stripes are slanted with respect to the applied perpendicular field direction (z). We know from the experimental results that as the applied field is increased, the \textbf{Q}-vector of the magnetic Bragg peak moves to a lower value, but maintains its orientation of 45$^{\circ}$ with respect to beam direction. This indicates that the majority domain expands but the stripe-domains maintains the 45$^{\circ}$ orientation. Step changes in multiples of 7 nm along L$_x$ imply that the x-component of the periodicity ($2\pi/Q_x$) changes in units of 7 nm.

Interestingly, this value matches with the exchange length ($L_{ex}$) of the studied Fe/Gd system. One of the ways to think about this behavior is that as the minority domains shrink (the horizontal region of the steps), there is a minimum distance between successive domain walls below which there cannot be a smooth deformation of the spin texture, as a result a sudden jump happens. In the theoretical section we will show that indeed by defining a term that signifies the ratio of spin kink to the spin chain, it is possible to predict jumps. At higher temperatures we observed an increase in the number of steps in the average-periodicity curves (see Fig.~\ref{Fig:QresolvedStaircase}(f and g) as a function of applied perpendicular field. This is due to the fact that thermal fluctuations aids in a faster transition from one step to the other as a result we obtain more number of steps at 236 K than 85 K even though the field range over which such steps occur is much higher at lower temperatures.

Existence of steps in solitonic systems with DMI has been observed experimentally and explained theoretically \cite{TogawaSoliton,PhysRevB.89.014419}. The presence of DMI introduces a topologically protected kink in the spin texture. The topological protection of the kink means that there is an energy cost to kink-annihilation. Different topological sectors have different energy which is the reason for step-like features. In contrast, in the Fe/Gd systems the dominant interactions are exchange, dipole, and anisotropy. This supports an achiral magnetic structure. So far there has being no theoretical studies of the step-like behaviour on dipole interaction dominant achiral spin-structures in an amorphous system.  In the theoretical model presented in the next section, we have mimicked the experimental conditions by investigating a one-dimensional dipolar mediated spin chain which is achiral in nature. We have numerically solved the LL equation of motion to understand the magnetization dynamics observed in the Fe/Gd system experiment. Based on our calculations we show that the origin of the step like behaviour under the application of an external perpendicular magnetic field could be explained by the spin dynamic behavior of an achiral spin chain.



\begin{figure}[!htbp]
\centering\includegraphics[width=2.85in]{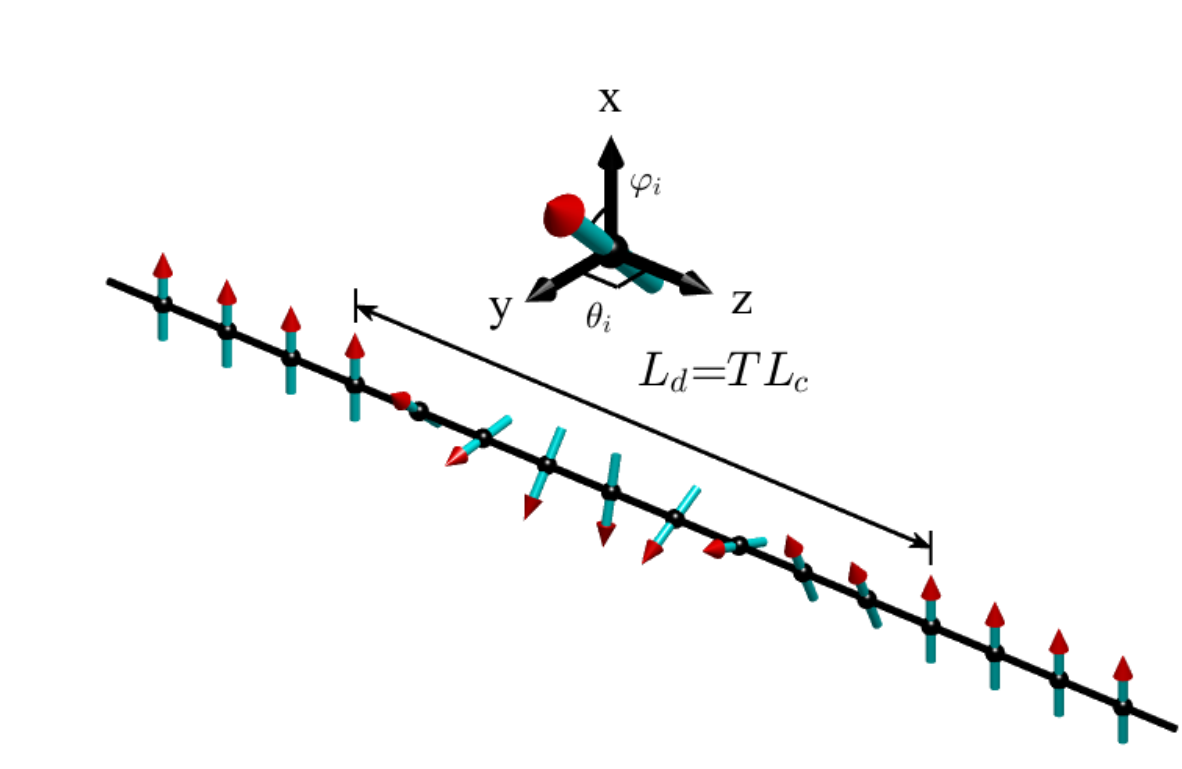}
\caption{{\bf Achiral spin chain arrangement.}~Achiral magnetic order is generated due to the competition between exchange interaction and dipolar interaction. The magnetic texture shown in the cartoon depicts an achiral spin arrangement where the spins rotate 180$^{\circ}$ out of plane (say in the positive screw direction, $+z$) and then rotate back to the original position (with an opposite negative screw rotation, $-z$). The spatial distance over which the achiral twist occurs can be captured by defining a local coefficient $T=L_{d}/L_{c}$, where $L_{d}$ is the length of the achiral domain area and $L_{c}$ is the chain length. The cartesian coordinate system shows the definition of the angle $\varphi_i$ and the angle $\theta$ used in the simulation.}
\label{fig:fig4}
\end{figure}

{\bf Model and theory}

The spin kinks caused by long range  dipolar interaction in the Fe/Gd system can be classified by a number $n$. In Fig.~\ref{fig:fig4} we show the local spin arrangement in a finite-size chain under zero applied magnetic field with fixed boundary condition on both ends. We consider a $N$-site 1D chain where spins interact with exchange interaction, dipolar interaction, anisotropy, the in-plane, and the out-of-plane (perpendicular) magnetic field. The spin on each site is parameterized as 
\begin{equation}
\label{eq:spin}
\boldsymbol{S}_i=(\sin\theta_i\cos\varphi_i,\sin\theta_i\sin\varphi_i,\cos\theta_i),
\end{equation}
where the site spin angle $\varphi_i=2\pi ni/\mathcal{N}$ and $\theta_i=\frac{\pi}{2}$. Here $i=0,1,2,...,\mathcal{N}$ where $N = \mathcal{N} + 1$. The kink sectors are classified by $n$ which indicate the number of domains existing in the chain. The Hamiltonian for our Fe/Gd system is
\begin{equation}
\label{eq:hamiltonian}
H=H_J+H_D+H_K+H_h,
\end{equation}
where the meaning and expression of each term is given by
\begin{equation}
\label{eq:exchange}
H_J=-J\sum\limits_{i~\in~\mathcal{N}}\boldsymbol{S}_i\cdot \boldsymbol{S}_{i+1}~~\text{(exchange)},
\end{equation}
\begin{equation}
\label{eq:dipolar}
H_D=D\sum\limits_{i,j~\in~\text{sc}}\boldsymbol{S}_i\cdot \boldsymbol{S}_j\Pi_{ij}~~\text{(dipolar interaction)},
\end{equation}
\begin{equation}
\label{eq:anisotropy}
H_K=-K_U\sum\limits_i(\boldsymbol{S}_i\cdot \boldsymbol{x})^2~~\text{(anisotropy)},
\end{equation}
\begin{equation}
\label{eq:field}
H_h=-g\mu_{B}H_x\sum\limits_iS^x_i-g\mu_{B}H_y\sum\limits_iS^y_i~~\text{(magnetic field)}.
\end{equation}

In the above $i$ either denotes the lattice site in the 1D chain or the location of a spin site inside a supercell (sc). The exchange interaction strength is given by $J >0$, the dipolar interaction coupling by $D$, the anisotropy by $K_U$, and the out- and in- plane magnetic field is given by $H_x$ and $H_y$, respectively. The symbol $g$ denotes the gyromagnetic ratio and the $\mu_B$ is the Bohr magneton. The $\Pi_{ij}$ in the dipolar interaction term is the Ewald coefficient which captures the long-range nature of the dipolar interaction. Using the angular representation of the spin $\boldsymbol{S}_i$ we can write the total energy as


\begin{equation}
\label{eq:H}
\begin{split}
\frac{H}{JS^2}&=-\sum\limits_{i \in N} \cos(\varphi_{i+1}-\varphi_i)+J_d\sum\limits_{i,j~\in~\text{sc}}\Pi_{ij}\cos(\varphi_i-\varphi_j)
\\&-K\sum\limits_i \cos^2\varphi_i-h_x\sum\limits_i\cos\varphi_i-h_y\sum\limits_i\sin\varphi_i,
\end{split}
\end{equation}
where we have now introduced the scaled variables $J_d=\frac{D}{J}$, $K=\frac{K_U}{J}$, $h_x=\frac{g\mu_BH_x}{J}$ and $h_y=\frac{g\mu_BH_y}{J}$. In all our figures we will report the scaled fields in milli-units, that is, $h_x = 1$ stands for $10^{-3}$ scaled field units. 



\begin{figure*}
\centering
\centering\includegraphics[width=6.5in]{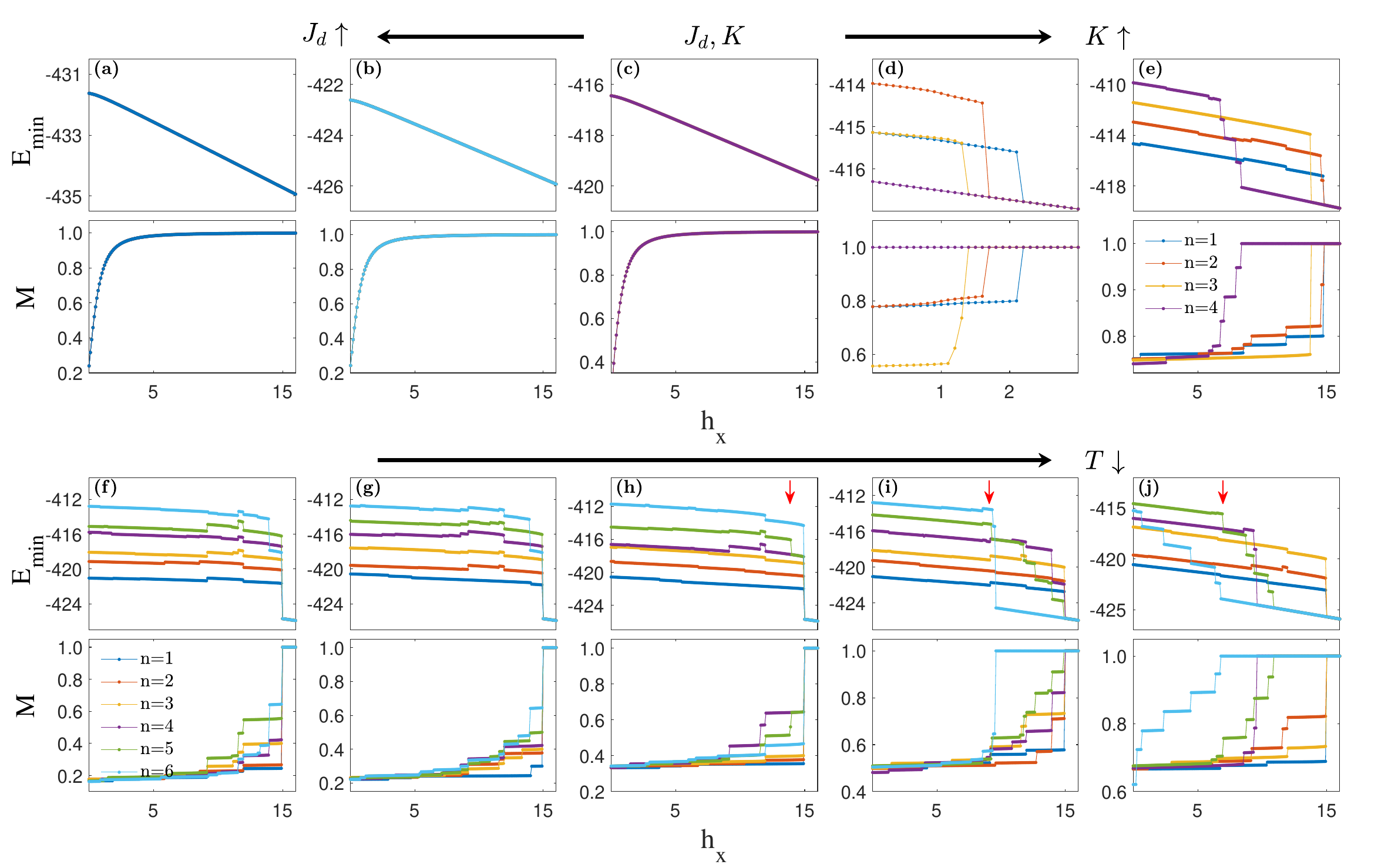}
\caption{\textbf{The energy and magnetization response for different $T$ values}. In the upper panel, dipolar parameter is $J_d=0.00934$ and the anisotropy parameter is $K=0.2$ (c). $J_d=0.00962,0.01004$ in (b) and (a) while anisotropy parameter $K$ remains the same as the one in (c). Anisotropy parameter $K=0.1,0.2$ in (d) and (e) while dipolar parameter remain the same as the one in (e). In the lower panel, $J_d=0.00962$ and $K=0.2$. The local coefficients from the left to the right are $T=\frac{7}{8},\frac{7}{9},\frac{2}{3},\frac{1}{2},\frac{1}{3}$,respectively. Red arrows in (h)-(j) represent the first jump for energy curve with $n=5$. The corresponding magnetic fields are $h_x=10.3,9.2,6.9$. All results are calculated with the number of sites $N=216$.}
\label{fig:fig5}
\end{figure*}

We implement the local achiral spin structure $N=216$ sites, shown in Fig.~\ref{fig:fig4}, to perform the LL simulation. To mimic the finite size of the experimental sample and to allow for the domains to grow and collapse as observed experimentally, we utilized an embedding trick to simulate the LL equations-of-motion (EOM). To capture experimentally realistic sample conditions, from a computational perspective, we introduced the concept of a local coefficient $T$. From a physical perspective, $T$ represents the ratio of the length of the achiral structure (which contains the twist sectors solely) over the length of the 1D chain. Thus, the achiral structure is embedded within a uniform ferromagnetic background spin texture. The computational embedding trick allows us to capture the spontaneous rearrangement of the twist sectors in the chain configuration, thereby simulating the growth and collapse of the achiral domain walls. Our numerical simulations indicate that the eventual fate of the twist sectors and subsequent realization of jumps (as observed experimentally) is a subtle balance between $J_d$, $K$, and $N$. We compute the minimum energy $E_{min}$ using Eq.~\eqref{eq:H}. The magnetization $M$ is calculated using
\begin{equation}
\label{eq:M}
M=\frac{1}{N}\sum\limits_{i=0}^{\mathcal{N}} \cos\varphi_i.
\end{equation} 

We present the energy and corresponding magnetization response of the local achiral state in Fig.~\ref{fig:fig5}. When the anisotropy is absent, we observe that the energy is degenerate for different twist sectors and no jumps are created by enhancing the dipolar interaction (see Fig.~\ref{fig:fig5}(a)-(c)). Moreover, it indicates that larger dipolar parameters induce a downshift in energy with no visible effects on the magnetization behavior in the local achiral state. In the presence of anisotropy, we keep the dipolar interaction constant and increase the $K$ parameter as shown in Fig.~\ref{fig:fig5}(c)-(e). We compute the LL dynamics on a chain of local achiral state with different anisotropy parameters. We find that upon enhancing anisotropy in the presence of a magnetic field, the energy degeneracy of the different twist sectors is broken with a simple upshift. With a relatively small $T=\frac{1}{4}$ and a strong enough anisotropy $K=0.2$, we observed jumps in both energy and magnetization in response to magnetic field as shown in Fig.~\ref{fig:fig5}(e).

In Fig.~\ref{fig:fig5}(f)-(j) we show our calculations of energy and magnetization response as $T$ is varied. With decreasing $T$, jumps begin to happen in energy response with higher twist sector. When $T>\frac{1}{2}$, jumps happen in energy curves with $n=6$ as shown in Fig.~\ref{fig:fig5}(f)-(h). However, jumps happen in energy curves with smaller twist sectors $n$ and lower magnetic field intensity $h_x$. In both Fig.~\ref{fig:fig5}(i) and Fig.~\ref{fig:fig5}(j), jumps happen when twist sector is $n\geqslant 4$. And the critical magnetic field intensity for the first jumps to happened decreases as $T$ decreases. It is found that energy response is more powerful to show the disappearance of kinks while the jumps in magnetization response might be caused by the position shifting of the kinks.

We have considered a chain with larger number of sites. When the number of sites is $N=432$ and the dipolar parameter $J_d=0.00916$, jumps can be observed in energy curve with twist sector $n=4$ and local coefficient $T=\frac{1}{4}$. However, no jumps can be observed with $T=\frac{1}{3}$ and $\frac{1}{2}$ (plots not shown). This behavior can also be seen in a system with $N=864$. The result that the decreasing $T$ contribute to the jumps, is also consistent with the $N=216$ system. Moreover, when the $J_d$ increases, more kinks are able to establish and more jumps are observed. Thus, we draw a conclusion that not just the declining local coefficient $T$, but also the rising dipolar parameter $J_d$ results in the jumps happening in energy curve with smaller twist sector $n$ and weaker magnetic field $h_x$. Note, that for particular parameters $J$, $D$, $K$ and the number of sites $N$, the value of the ground state energy remains almost the same when the local coefficient $T$ changes.

{\bf DISCUSSION}


In this work we have shown experimentally that in an amorphous and centrosymmetric Fe/Gd magnetic thin film that exhibits stripe and skyrmion lattice phases, the stripe-domain periodicity changes in steps because of the abrupt disappearance of stripe-domains. This result is interesting in itself because similar to a DMI based solitonic system, exchange-dipole mediated Fe/Gd system also shows similar step-like behavior even though a sole global chirality is absent in the system. Since the presence of a global DMI can be ruled out in the Fe/Gd systems~\cite{JamesLee,Sergio}, we can conclude that the predicted spin twists are formed due to the competition between exchange and dipolar interactions, and the spin twist sectors can be smoothly transformed to the uniform phase by any number of finite deformations. 

Intuitively, due to the achiral spin texture of the domains, as the magnetic field is increased, the minority domains start to shrink, resulting in two "like-domains" to come closer. The minimum distance between the two domains is guided by two spin-kinks on either side which should be equivalent to the length of two domain walls. Using the well known formula $l_w=\sqrt{J/K}$, where $l_w$ is the domain wall width, the domain wall width for Fe/Gd comes out $\approx$ 3.2 nm, twice of which is 6.4 nm, which is in close agreement to the experimental value of 7 nm. Thus the minimum distance between the two like-domains comes out to be equivalent to the exchange length ($L_{ex}$) of the system from our experimental study. The above explanation also points to the existence of a ``global" and ``local" length scales in the system which will give rise to two energy scales. It is these competing energy scales that give rise to steps. Our system is reminiscent of the case of modulated periodicities.

Our 1D model suggests the existence of discrete magnetization steps in the REXS experiment results from magnetic spin textures exhibiting achiral spin-twist characteristics. Recently, a transport and micromagnetic study~\cite{SM2022} of a patterned Fe/Gd specimen, with same material composition as the one addressed in this work, suggests the domain walls of the stripe domains undergo a local chirality spin rearrangement from chiral to achiral, under similar applied perpendicular field conditions, which results in stripe domains exhibiting achiral spin-twist characteristics as the one addressed in the 1D model of this paper. Although Ref.~\onlinecite{SM2022} addresses the formation mechanism of skyrmion lattice phases in the Fe/Gd system specimen, close inspection of the field-dependent micromagnetic vector magnetization $(m_x,m_y,m_z)$ evolution, in their work, shows the magnetization components attributed to the domain wall $(m_x,m_y)$ undergoes abrupt/sharp changes as the perpendicular field is swept from zero-field towards magnetic saturation which are likely correlated to the collapse and local rearrangement of magnetic spin textures with achiral spin-twist characteristics.

We take the achiral nature of the stripe spin structure as an important point in our theoretical development and show that the magnetization steps can indeed be observed in a dipolar magnet with net global achirality. The variations and interplay of the length scales is captured in the parameter $T$. Analysis of the energy expression with different values of $T$ suggests that in an achiral spin arrangement staircase structure can be observed only under certain specific ratios of 1D spin to spin-twist length scale. Although simplistic, our LL calculations using local achiral spin structure shown in Fig.~\ref{fig:fig4} is able to capture the essential feature that the system has jumps in response to an external magnetic field. The jumps happen only when anisotropy is present. Absence of anisotropy leads to a degeneracy of energy response for different twist sectors, meaning absence of jumps in the system. Our study provides evidence and further impetus to study magnetic spin textures in a centrosymmetric magnetic, both from an experimental and theoretical viewpoint.




{\bf METHODS}

\textbf{Sample details}

The samples studied were nominally [Gd (0.4 nm)/Fe(0.34 nm)] $\times$ 80 multilayers deposited using DC magnetron sputtering with 20 nm Ta seed and capping layers. The samples were deposited on 50-nm or 200-nm thick Si3N4 membranes to allow for transmission RSXS experiments, respectively. Non-resonant 12 keV x-ray diffraction indicated strong intermixing of Gd and Fe layers thereby forming an amorphous structure rather than a multilayer. Given the centrosymmetric nature of the Fe/Gd system, negligible Dzyaloshinskii–Moriya exchange interactions are expected~\cite{JamesLee,Singh}.

\textbf{Experimental details}

The coherent X-ray magnetic scattering measurements were performed at beamline 12.0.2.2 of the Advanced Light Source, LBNL. The incident beam was tuned to the Fe L$_3$ edge  (707 eV). Transverse coherence of the X-ray beam was established by inserting a 10 µm pinhole in the beampath before the sample. The scattering experiment was done in the transmission geometry at temperatures ranging from 40 K to 300 K as a function the perpendicular magnetic field from 0 mT to 500 mT. (Fig.~\ref{Fig:scattering}(a)). The sample was subjected to the following initial magnetic field protocol. First the field was raised to 500 mT, then lowered to -500 mT and finally to zero before taking the measurements. The field ramp rate for the first two legs is 13 mT/sec, while the final drop of field from -500 mT to 0 mT took place at a rate of 380 mT/sec. We start our measurement at this zero-field condition and proceed to measure the diffraction signal as a function of applied magnetic field at a constant rate of 1.575 mT/s. A Charge Coupled Device (CCD) camera placed at about 0.5 m downstream of the sample was used to record the scattered intensity patterns.

\textbf{Ewald method} 

In the Hamiltonian calculation, Ewald summation is applied, which is given by
\begin{equation}
\label{eq:ewald}
\begin{split}
\Pi_{ij}&=\sqrt{\frac{2}{\pi}}\frac{1}{3\sigma^3}\sum\limits_{n}e^{-\frac{\left|r_{ij}- \mathsf{n} L\right|^3}{2\sigma^2}}+\frac{4\pi}{\Omega}\sum\limits_{k\neq 0}e^{-\frac{\sigma^2k^2}{2}}\cos(kr_{ij})\\&-\sqrt{\frac{2}{\pi}}\frac{1}{3\sigma^3}\delta_{ij},
\end{split}
\end{equation}
where $r_{ij}$ represents the distance between two spin sites, $L$ is the size of the supercell, $n$ is the supercell label, $\sigma$ is the real-space cut-off, $k$ is the momentum space label, and $\Omega$ is the volume of the supercell which in our case is equal to $L$. The Ewald parameter will be redefined as $\Pi_{ij} =\Pi_{|i-j|}\equiv\Pi_{m}$, where the symbol $m$ tracks the number of Ewald parameters for the specific supercell size choice. The values of $\Pi_{m}$ are shown in Table.~\ref{tab:table1}.
\begin{table}[t]
\centering
\caption{Ewald parameter symbols and the corresponding Ewald parameter values given by Eq.~\eqref{eq:ewald}.} 
\label{tab:table1}
\begin{tabular}{c|c}
\hline\hline
\textbf{Ewald parameter symbol} & \textbf{Ewald parameter value} \\ \hline
$\Pi_1$ & 2.00 \\ 
$\Pi_2$ & 1.72 \\ 
$\Pi_3$ & 1.32 \\ 
$\Pi_4$ & 0.85 \\ 
$\Pi_5$ & 0.38 \\ 
\hline\hline
\end{tabular}   
\centering
\end{table}

\textbf{Landau-Lifshitz (LL) equation of motion}

We perform a LL EOM spin dynamics calculation on Eq.~\ref{eq:hamiltonian}. We obtained an iterative equation which can be used to calculate the angle $\varphi_i$ of each spin on the chain. Based on our computations, we are able to generate a stabilized spin order along the chain. Next, we computed the twist angle $\Delta\varphi$ of the ground state in the absence of an external magnetic field using the energy minimization condition for the supercell given by the expression
\begin{equation}
\label{eq:GS}
\frac{E}{JS^2}=-\cos\Delta\varphi+J_d\sum\limits_{m=1}^{L-1}(L-m)\cos(m\Delta\varphi)\Pi_m.
\end{equation}

The angle $\varphi_i$ was analyzed to obtain the relationship between the range of $J_d$ and the number of kinks for a given lattice size of site $N$. The relationship between the number of sites $N$, dipolar parameter $J_d$ and the maximal sector $n_{max}$ is shown in Table.~\ref{tab:table2}. Note that to perform the calculation, one needs to choose a supercell size that stabilizes the ground state and ensures that there will be minimal to no numerical oscillations in the computed result due to convergence issues. We found that $L=6$ is the optimal supercell size which yields numerically stable results for our LL analysis. Using the numerically stable data, we computed the minimum energy $E_{min}$ (scaled relative to $J S^2$) and magnetization $M$ (scaled relative to $S$). To compare our numerical results with the experimental setup of the Fe/Gd system, we need to mimic the experimental conditions. Therefore, all the results are calculated by applying a tiny in-plane field $h_y$.

\begin{table}[t]
\centering
\caption{Maximum allowed number of kink sectors $n_{max}$ under different scaled dipolar parameter $J_d$ for a given number of lattice sites $N$. 
We introduce a sector $n$ denoting the number of twisted magnetic structures in a achiral lattice. The system relaxes to the ground state with a maximal sector value $n_{max}=\left[\frac{N\Delta\varphi}{2\pi}\right]$, where the square bracket implies the flooring function.} 
\label{tab:table2}
\begin{tabular}{ccc}
\hline\hline  \textbf{Number of sites} & \textbf{Dipolar $J_d$} & \textbf{Sector $n_{max}$} \\ \hline
\multirow{3}{*}{216} & 0.00934 & 4  \\ 
                     & 0.00962 & 6  \\ 
                     & 0.01004 & 8  \\ \hline
\multirow{3}{*}{432} & 0.00916 & 4  \\ 
                     & 0.00934 & 8  \\ 
                     & 0.00962 & 12 \\ \hline  
\multirow{3}{*}{864} & 0.00916 & 8  \\ 
                     & 0.00934 & 17 \\ 
                     & 0.00962 & 25 \\ \hline\hline  
\end{tabular}   
\centering
\end{table} The two angular variable EOMs are given by (for $\hbar=1$)
\begin{equation}
S\sin\theta_i\partial_t\theta_i=\frac{\partial H}{\partial\varphi_i},~~~S\sin\theta_i\partial_t\varphi_i=-\frac{\partial H}{\partial\theta_i},
\label{eq:LL}
\end{equation}
where only the first equation is required because the angle $\theta$ is held constant. Using Eq.~\eqref{eq:LL} we can obtain the following expressions
\begin{equation}
\begin{split}
&0=\frac{1}{2}\left[\sin(\varphi_i-\varphi_{i-1})-\sin(\varphi_{i+1}-\varphi_i)\right]
\\&+J_d\sum\limits_{j~\in~ \text{sc}}\left[\sin(\varphi_{i+|i-j|}-\varphi_i)-\sin(\varphi_i-\varphi_{i-|i-j|})\right]\Pi_{ij}
\\&+2K\sin\varphi_i \cos\varphi_i+h_x\sin\varphi_i-h_y\cos\varphi_i.
\end{split}
\label{eq:stable}
\end{equation}
The above can be split further into a form convenient for a numerical iterative self-consistent approach to solve for the angle $\phi_i$. Hence, we write
\begin{eqnarray}
\label{eq:a}
A_i&=\frac{1}{2}(\sin\varphi_{i+1}+\sin\varphi_{i-1})-J_d\sum\limits_{j~\in ~sc}(\sin\varphi_{i+|i-j|}\nonumber
\\&+\sin\varphi_{i-|i-j|})\Pi_{ij}-K\sin\varphi_i+h_y,
\end{eqnarray}
\begin{eqnarray}
\label{eq:b}
B_i&=\frac{1}{2}(\cos\varphi_{i+1}+\cos\varphi_{i-1})-J_d\sum\limits_{j~\in ~sc}(\cos\varphi_{i+|i-j|}\nonumber
\\&+\cos\varphi_{i-|i-j|})\Pi_{ij}+K\cos\varphi_i+h_x,
\end{eqnarray}
with the site angle $\varphi_i$ is defined as 
\begin{equation}
\label{eq:cossine}
\sin\varphi_i=\frac{A_i}{\sqrt{A_i^2+B_i^2}},~~~\cos\varphi_i=\frac{B_i}{\sqrt{A_i^2+B_i^2}}.
\end{equation}

The LL equation is solved with the boundary condition $\varphi_0=0$ and $\varphi_{\mathcal{N}}=2\pi n$. To obtain the final spin structure, we need to do the following things. First, we set the spin structure as a local achiral structure in which ratio of the length of the domain area over the chain length equals to $T$ (see Fig.~\ref{fig:fig4}). Second, we setup the dipolar and anisotropy parameters and run the LL simulation program to compute the stabilized spin structure. Finally, we change the magnetic field to compute the system’s magnetization and energy response to the magnetic field.

{\bf ACKNOWLEDGEMENTS}

This work used resources of the Advanced Light Source, which is a DOE Office of Science User Facility under contract no. DE-AC02-05CH11231 (X-ray scattering). The research at UCSD was supported by the National Science Foundation, Division of Materials Research (Award \#: 2105401) (Materials synthesis and characterization). The theory modeling and LL simulation was performed by J. L., D. X. Y., and T.D. J. L. and D. X. Y were supported by NKRDPC-2022YFA1402802, NKRDPC-2018YFA0306001, NSFC-92165204, and NSFC-11974432. T. D. acknowledges hospitality of KITP, where a part of the theory modeling and LL simulation research was completed, which was supported in part by the National Science Foundation under Grant No. NSF PHY-1748958. T.D. acknowledges valuable and insightful discussions on domain dynamics with Ulrich R\"{o}{\ss}ler.

{\bf DATA AVAILABILITY}

The data sets generated during the current study are available from the corresponding authors upon reasonable request.

{\bf CODE AVAILABILITY}

The codes utilized during the current study are available from the corresponding authors upon reasonable request.

{\bf COMPETING INTERESTS}

The authors declare no competing financial or non-financial Interests.

{\bf AUTHOR CONTRIBUTIONS} 

S.R. and A.S. conceived the experiment. A.S. and S.R. performed X-ray experiments. A.S., S.M. S.R. S.D.K. and P.F., analyzed the data and discussed experimental interpretation. S.A.M. and E.E.F. synthesized samples and performed magnetic characterization. The theory was conceived by T.D., J. L., and D. X.Y. J. L. performed the calculations. T.D and D.X.Y. checked the calculations. A. S., J. L., S.M., D. X. Y, T. D., and S. R. wrote the manuscript in consultation with the other authors. 

\bibliographystyle{apsrev4-1}
\bibliography{main}

{\bf FIGURE AND TABLE LEGENDS} 
\begin{itemize}
\item Figure 1: {\bf Experimental set-up and magnetic phases.} (a) Schematic of the coherent magnetic X-ray scattering geometry. 
(b) Real space (top panel) and reciprocal space (lower panel) of the different magnetic phases present in a Fe/Gd system. The scattering images were taken at H = 0mT; T = 85K (disorder stripes), H = 0mT;  T = 225K (order stripes) and H = 190 mT; T = 239K (skyrmions). A small residual in-plane field is present during ramping down of field from saturation to zero. (c) Variation of the 1st and 2nd order magnetic diffraction peak with field at 239K. Inset image shows the appearance of both 1st and 2nd order diffraction peaks.

\item Figure 2: {\bf Evolution of stripe diffraction peak, periodicity and correlation with field.} (a) Plot of the q-vector of the satellite peaks as a function of the applied out-of-plane (OOP) magnetic field at 230 K as the system transitions from magnetic stripe phase to skyrmion phase. Dotted arrows indicate the Q-value of the Bragg peak positions (purple symbol) starting from at different fields. (b) Evolution of the stripe-periodicity with field at various temperatures showing discrete steps like feature. (c) Correlation coefficient values with respect to the remanent state (0 mT) for increasing magnetic field at 85K.

\item Figure 3: {\bf Stripe orientation and staircase-like behaviour.} (a) A typical scattering pattern of the stripe lattice along with the projection of the in-plane Q-vectors. (b) Enlarged image of the stripe-diffraction spot in Q-space. (c) Schematic real space view of stripe-domain orientation according to scattering image of Fig.~\ref{Fig:staircase}(a), where blue circles with dot resemble the spin along the field direction while the red small circles with cross resemble the spins opposite to the field direction and L= L${_x}{^2}$+L${_y}{^2}$ corresponds to the periodicity of the stripe domains. Plot of the evolution of (d) L$_y$ (= 2$\pi$/Q$_y$) at 85 K and (e-g) L$_x$ (= 2$\pi$/Q$_x$) as a function of the applied magnetic field at T = 85 K, 183 K and 236 K.

\item Figure 4: {\bf Achiral spin chain arrangement.}~Achiral magnetic order is generated due to the competition between exchange interaction and dipolar interaction. The magnetic texture shown in the cartoon depicts an achiral spin arrangement where the spins rotate 180$^{\circ}$ out of plane (say in the positive screw direction, $+z$) and then rotate back to the original position (with an opposite negative screw rotation, $-z$). The spatial distance over which the achiral twist occurs can be captured by defining a local coefficient $T=L_{d}/L_{c}$, where $L_{d}$ is the length of the achiral domain area and $L_{c}$ is the chain length. The cartesian coordinate system shows the definition of the angle $\varphi_i$ and the angle $\theta$ used in the simulation.

\item Figure 5: \textbf{The energy and magnetization response for different $T$ values}. In the upper panel, dipolar parameter is $J_d=0.00934$ and the anisotropy parameter is $K=0.2$ (c). $J_d=0.00962,0.01004$ in (b) and (a) while anisotropy parameter $K$ remains the same as the one in (c). Anisotropy parameter $K=0.1,0.2$ in (d) and (e) while dipolar parameter remain the same as the one in (e). In the lower panel, $J_d=0.00962$ and $K=0.2$. The local coefficients from the left to the right are $T=\frac{7}{8},\frac{7}{9},\frac{2}{3},\frac{1}{2},\frac{1}{3}$,respectively. Red arrows in (h)-(j) represent the first jump for energy curve with $n=5$. The corresponding magnetic fields are $h_x=10.3,9.2,6.9$. All results are calculated with the number of sites $N=216$.

\item Table I: Ewald parameter symbols nad the corresponding Ewald parameter values given by Eq.~\eqref{eq:ewald}.

\item Table II: Maximum allowed number of kink sectors $n_{max}$ under different scaled dipolar parameter $J_d$ for a given number of lattice sites $N$. We introduce a sector $n$ denoting the number of twisted magnetic structures in a achiral lattice. The system relaxes to the ground state with a maximal sector value $n_{max}=\left[\frac{N\Delta\varphi}{2\pi}\right]$, where the square bracket implies the flooring function.

\end{itemize}

\end{document}